\documentclass[10pt,twocolumn,letterpaper]{article}

\usepackage{cvpr}      

\usepackage{graphicx}
\usepackage{amsmath}
\usepackage{amssymb}
\usepackage{booktabs}
\usepackage{multirow} 

\usepackage[pagebackref,breaklinks,colorlinks]{hyperref}

\usepackage[capitalize]{cleveref}
\crefname{section}{Sec.}{Secs.}
\Crefname{section}{Section}{Sections}
\Crefname{table}{Table}{Tables}
\crefname{table}{Tab.}{Tabs.}

\begin{document}

\title{Unveiling the Era of Spatial Computing}

\author{HanZhong Cao\\
Peking University\\
{\tt\small 2200012929@stu.pku.edu.cn}}

\maketitle

\begin{abstract}
    The evolution of User Interfaces (UIs) marks a significant transition from traditional command-line interfaces to more intuitive graphical and touch-based interfaces, largely driven by the emergence of personal computing devices. The advent of spatial computing and Extended Reality (XR) technologies further pushes the boundaries, promising a fusion of physical and digital realms through interactive environments. This paper delves into the progression from All Realities technologies—encompassing Augmented Reality (AR), Virtual Reality (VR), and Mediated Reality—to spatial computing, highlighting their conceptual differences and applications. We explore enabling technologies such as Artificial Intelligence (AI), the Internet of Things (IoT), 5G, cloud and edge computing, and blockchain that underpin the development of spatial computing. We further scrutinize the initial forays into commercial spatial computing devices, with a focus on Apple's Vision Pro, evaluating its technological advancements alongside the challenges it faces. Through this examination, we aim to provide insights into the potential of spatial computing to revolutionize our interaction with digital information and the physical world.
\end{abstract}

\section{Introduction}
\label{sec:intro}

\begin{figure*}[h]
    \centering
    \includegraphics[width=0.8\textwidth]{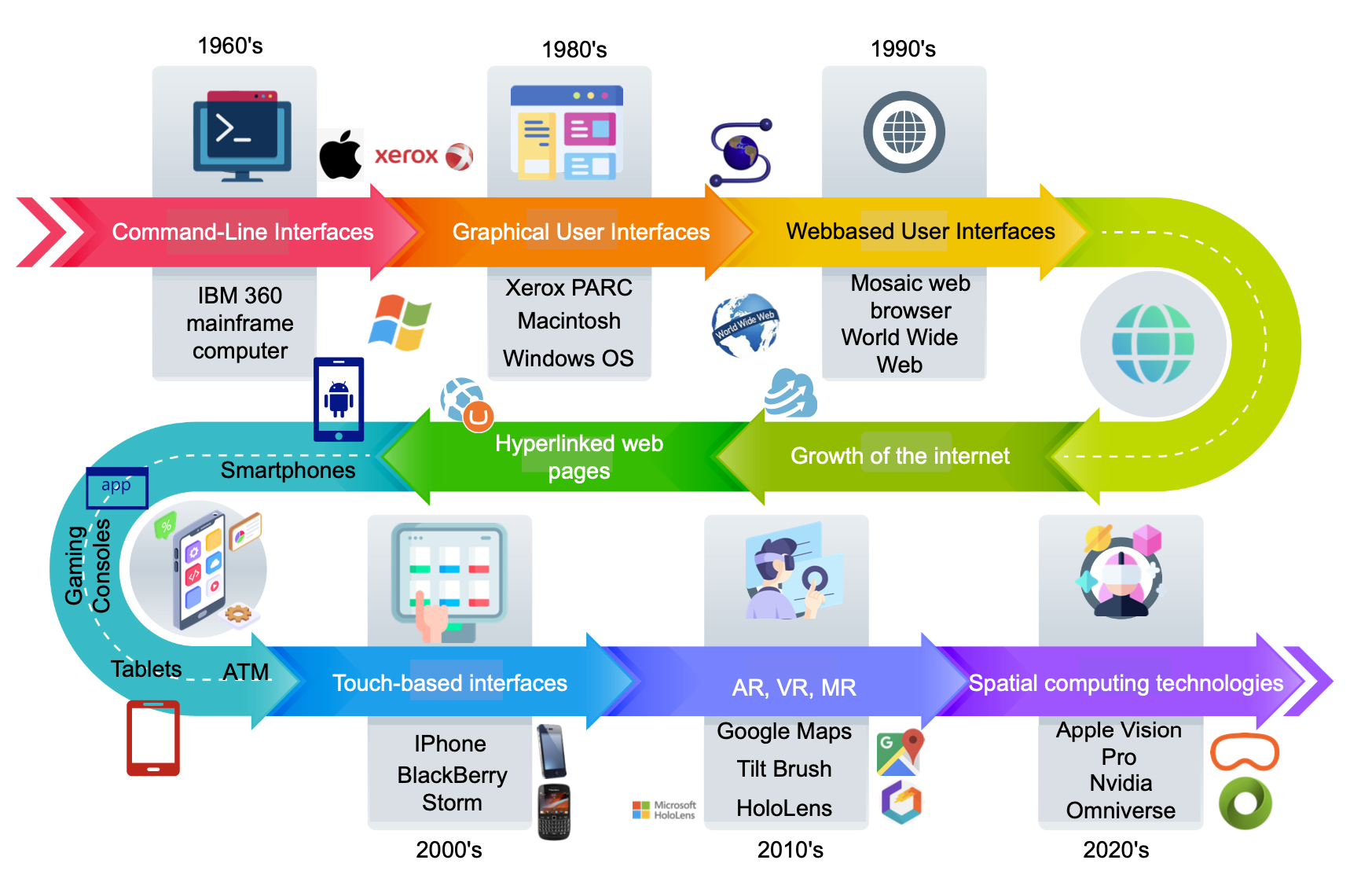}
    \caption{Road Map of User Interfaces,imported by \cite{yenduri2024spatial}}
    \label{figure:2}
\end{figure*}

Since the advent of the information age, user interfaces (UIs) have experienced a myriad of reforms and innovations. The evolution from Command Line Interfaces (CLIs) to Graphical User Interfaces (GUIs) has made computers increasingly user-friendly. With the rise of smartphones and tablets, touch-based interfaces have become mainstream for Internet browsing. The latest advancements are in Extended Reality (XR) technologies and spatial computing, heralding a new era of user interfaces.The road map of Use interfaces is visualized in \ref{figure:2}.

Spatial computing allows the collection of dynamic physical reality data to construct a digital space. This space, though low in dimension, is crafted to be rich enough for human perception. It encompasses a variety of physical data, including spatial location, audio information, peripheral vision, distance metrics, eye and body movements, and even subjective judgements and attitudes on a broader scale. The resulting digital environment can take various forms and effects, contingent upon the specific concepts it serves. For instance, advances in XR technologies like Mixed Reality (MR), Augmented Reality (AR), and Virtual Reality (VR) offer differentiated levels of virtual immersion. Moreover, the latest iteration of internet-based applications, such as the metaverse, integrates social and economic systems within virtual worlds, striving for multi-technology convergence\cite{ning2021survey}. Spatial computing aims to achieve a higher consistency and a reduced sense of discord in virtual synthetic human experiences, serving as an "external eye" for users.

\section{All realities}

Technologies encompassing all realities serve as the foundational precursor to spatial computing, demonstrating transformative impacts across various fields and heralding a promising future for spatial computing. This discussion begins with an overview of the concepts underpinning all realities, followed by an examination of their applications and the enhancements that future spatial computing technologies may afford.

\begin{figure*}[h]
    \centering
    \includegraphics[width=0.8\textwidth]{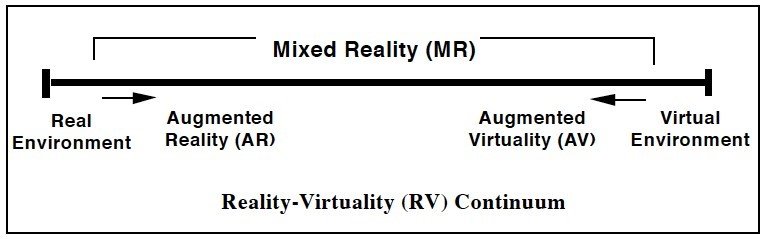}
    \caption{mixed reality continuum}
    \label{figure:1}    
\end{figure*}

X-reality encompasses at least three definitions, determined by their relationship within the mixed reality spectrum\cite{mann2018reality}. A simplified way to understand this is by referring to the mixed reality continuum represented in \ref{figure:1}, which illustrates the degree of virtualization\cite{PaulMilgram}. Adjusting a one-dimensional “fader” shifts the experience between different reality technologies like AR and VR. Moreover, by examining the extent to which reality can be augmented, we introduce the concept of mediated reality, which directly alters our perceived reality\cite{mann2018reality}. The subsequent sections detail the applications of these technologies.

\textbf{AR (Augmented Reality)}: AR enhances human-computer interaction by superimposing computer-generated 3D content onto the real world, aiming to enrich multiple sensory modalities and thereby augment the perception of reality. AR finds direct application in visualizing geographic data \cite{ijgi10090593}, measuring and depicting damage in post-disaster assessments\cite{doi:10.1061/(ASCE)0887-3801(2007)21:5(303)}, and enhancing virtual travel experiences by displaying virtual models across various environments\cite{ijgi7120479}.

\textbf{VR (Virtual Reality)}: VR offers a simulated experience that immerses users in a virtual environment, aiming to isolate them from real-world surroundings. This isolation facilitates interactive simulations for rapid assessment and iterative development of novel design concepts unconstrained by financial or security limitations\cite{DBLP:journals/corr/abs-2402-15695}. A notable application is the European Space Agency's use of VR in designing the Argonaut lunar lander, allowing for immersive experiences that elicit genuine and effective user feedback.

\textbf{Mediated Reality}: This extends beyond AR to either modify or mediate our perceived reality. It can be intentionally mediated, such as through invert-vision glasses, or unintentionally mediated, often used in treating PTSD with head-mounted displays.

Spatial computing, in essence, seeks to merge the best aspects of AR and VR while striving for the authenticity afforded by mediated reality. It operates on the principle of understanding and interacting with the physical world through digital means, fulfilling all related demands. This approach facilitates a seamless integration of digital and physical elements. Below is Table\ref{tab:1}, which compares the aforementioned concepts.

\begin{table*}[h]
\centering
\begin{tabular}{|c|p{3cm}|p{3cm}|p{3cm}|p{3cm}|}
\hline 
\textbf{Parameters} & Augmented Reality & Virtual Reality & Mediated Reality & Spatial Computing\\
\hline
\multirow{4.5}{*}{Definition} & Overlays virtual content onto the real world, enhancing perception of reality & Creates a fully immersive simulated environment, replacing the real world & Adding to, subtracting information from, or otherwise manipulating one's perception of reality & Blends the real and virtual worlds, enabling interaction and perception in a spatial context \\
\hline
Hardware & Smartphones tablets AR glasses heads-up displays & VR headsets HMDs immersive devices & a wearable computer or hand-held device & smart headsets devices \\
\hline
Interaction & Gesture-based, touch-based, voice commands & Controller-based, hand tracking & tracking system & Gesture-based, touch-based, voice commands, tracking systems \\
\hline
Immersion & Partial immersion, overlaying virtual content on the real world & Full immersion, replacing the real world with a virtual environment & Full immersion, modifying reality as fatigue-free as possible & Varies depending on the application and hardware used \\
\hline
Use Cases & Gaming, education, training, navigation, entertainment & Gaming, simulations, training, entertainment & seeing aid, design & Architecture, interior design, industrial applications, collaborative workspace \\
\hline
\end{tabular}
\caption{Immersive Technologies Comparison Matrix ,modified by 
\label{tab:1}
\cite{yenduri2024spatial}}
\end{table*}

\section{Enabling Technologies of Spatial Computing}

In this section ,we will shed a light on the versatile application potential of spatial computing.That we will break it into five aspects to delineate it.

\textbf{Artificial Intelligence}Combining the physical environment and digital information of virtual objects requires artificial intelligence to understand the scene, identify and track objects in the virtual environment, recognize gestures, detect interactions between objects, understand and handle object occlusion, and map them with the real-time spatial layout. Artificial Intelligence can be perceived as the core technology and fundamental concept of spatial computing, as it can evolve from every progress made in computer vision tasks.

\textbf{IOT} The Internet of Things (IoT) is the pivotal concept of communication between different kinds of devices, including those with sensors, processing ability, and display capabilities. Spatial computing requires various sensors and cameras to collect data from the physical environment. By processing this data, spatial computing devices can also serve as IoT network center processors, helping humans better control the IoT and revolutionizing how we interact with technology.

\textbf{5G and beyond}The high demand for low latency in spatial computing cannot be handled without 5G and beyond technologies. By facilitating a deeper connection between physical and digital spaces, 5G and beyond technologies enable seamless transitions between virtual reality and the real world. Only the advent of 6G technologies will bring immersive experiences to social contexts, opening up possibilities for using spatial computing in education, entertainment, and various other industries highly dependent on networks.

\textbf{Cloud/Edge Computing} For portability, spatial computing may rely on both cloud and edge computing as part of a distributed architecture to solve heavy computing tasks like rendering 3D models. The ability to manage and store data also plays a crucial role in spatial computing. Distributed or edge computing can save bandwidth and reduce costs when spatial computing becomes ubiquitous.

\textbf{Blockchain}The security of a similar concept like the metaverse is mainly protected by blockchain technology, which also has value when spatial computing aims to provide internet digital worlds. Due to its decentralized nature, people can use spatial computing devices more privately without worrying about third-party services stealing their biometric information.

\section{The first products:Apple vision pro and its challenges}

In 2023, spatial computing made significant strides with the launch of its first notable product, Apple's Vision Pro, a headgear endowed with AR, VR, and MR capabilities\cite{applevisionpro}. Devices like the Apple Vision Pro chart user surroundings using an array of sensors and cameras, employing infrared cameras to establish a sophistical high-performance eye-tracking system\cite{applecom}. Additionally, it uses a pair of high-resolution cameras for collecting reality data at a high refresh rate, enabling the rendering of dynamic spatial digital images. The integration of a LiDAR Scanner and a TrueDepth camera in the Vision Pro works in tandem to generate a composite 3D map. With these technologies, spatial computing can overlay virtual objects into the real world seamlessly, provided a spatial 3D map of the area is obtained and there is an understanding of the objects within it.

However, the Apple Vision Pro faces several challenges. The first significant hurdle is its high cost, attributed not only to the substantial investment in its development phase but also to the intrinsic expenses related to its hardware and software requirements\cite{visionprochallenge}. Additionally, discomfort during extended use poses a second major challenge\cite{visionprochallenge2}. There are also reports indicating a slight discrepancy between the real position of the user's fingers and their representation in digital spaces, which could prove detrimental for precision tasks. These issues underscore the fact that spatial computing still has considerable ground to cover before it can be fully embraced by the public.

\section{Summary}

The paper begins by charting the evolution of user interfaces, culminating in the conceptualization and application of spatial computing—a technology that marries the physical and digital through the perception and interaction within a constructed digital space. Various All Realities technologies such as AR, VR, and mediated reality are discussed, including their specific applications and the transition towards a more integrated experience provided by spatial computing. The enabling technologies section outlines AI, IoT, 5G, cloud and edge computing, and blockchain as essential components that drive spatial computing capabilities. A case study on Apple's Vision Pro is presented, showcasing an example of spatial computing's application and the challenges faced, such as cost, user comfort, and integration accuracy. The paper highlights the transformative potential of spatial computing across numerous domains, despite the nascent challenges that currently limit widespread adoption.
must include your signed IEEE copyright release form when you 


\appendix

{\small
\begin{thebibliography}{10}\itemsep=-1pt

\bibitem{applecom}
apple.
\newblock https://www.apple.com/apple-vision-pro/.
\newblock Technical report, apple, 2024.

\bibitem{DBLP:journals/corr/abs-2402-15695}
Leonie Bensch, Andrea E.~M. Casini, Aidan Cowley, Florian Dufresne, Enrico Guerra, Paul de Medeiros, Tommy Nilsson, Flavie Rometsch, Andreas Treuer, and Anna Vock.
\newblock Applied user research in virtual reality: Tools, methods, and challenges.
\newblock {\em CoRR}, abs/2402.15695, 2024.

\bibitem{visionprochallenge2}
Alex Blake.
\newblock Report claims vision pro sales have ‘fallen sharply’ as apple struggles to whip up demand.
\newblock Technical report.

\bibitem{ijgi10090593}
Kejia Huang, Chenliang Wang, Shaohua Wang, Runying Liu, Guoxiong Chen, and Xianglong Li.
\newblock An efficient, platform-independent map rendering framework for mobile augmented reality.
\newblock {\em ISPRS International Journal of Geo-Information}, 10(9), 2021.

\bibitem{doi:10.1061/(ASCE)0887-3801(2007)21:5(303)}
Vineet~R. Kamat and Sherif El-Tawil.
\newblock Evaluation of augmented reality for rapid assessment of earthquake-induced building damage.
\newblock {\em Journal of Computing in Civil Engineering}, 21(5):303--310, 2007.

\bibitem{mann2018reality}
Steve Mann, Tom Furness, Yu Yuan, Jay Iorio, and Zixin Wang.
\newblock All reality: Virtual, augmented, mixed (x), mediated (x,y), and multimediated reality, 2018.

\bibitem{PaulMilgram}
Paul Milgram, Haruo Takemura, Akira Utsumi, and Fumio Kishino.
\newblock Augmented reality: A class of displays on the reality-virtuality continuum.
\newblock {\em Telemanipulator and Telepresence Technologies}, 2351, 01 1994.

\bibitem{ning2021survey}
Huansheng Ning, Hang Wang, Yujia Lin, Wenxi Wang, Sahraoui Dhelim, Fadi Farha, Jianguo Ding, and Mahmoud Daneshmand.
\newblock A survey on metaverse: the state-of-the-art, technologies, applications, and challenges, 2021.

\bibitem{visionprochallenge}
Reuters.
\newblock Apple scales back vision pro production plans on design challenges, financial times reports.
\newblock Technical report.

\bibitem{ijgi7120479}
Piotr Siekański, Jakub Michoński, Eryk Bunsch, and Robert Sitnik.
\newblock Catcha: Real-time camera tracking method for augmented reality applications in cultural heritage interiors.
\newblock {\em ISPRS International Journal of Geo-Information}, 7(12), 2018.

\bibitem{applevisionpro}
Ong J. Masalkhi M. et~al. Waisberg, E.
\newblock The future of ophthalmology and vision science with the apple vision pro.eye 38, 242–243.
\newblock 2024.

\bibitem{yenduri2024spatial}
Gokul Yenduri, Ramalingam M, Praveen Kumar~Reddy Maddikunta, Thippa~Reddy Gadekallu, Rutvij~H Jhaveri, Ajay Bandi, Junxin Chen, Wei Wang, Adarsh~Arunkumar Shirawalmath, Raghav Ravishankar, and Weizheng Wang.
\newblock Spatial computing: Concept, applications, challenges and future directions, 2024.

\end{thebibliography}

}
\end{document}